\documentstyle[12pt]{article}\textwidth=6.5in \oddsidemargin=0in
\begin{document}

\newcommand{\bq}{\begin{equation}} \newcommand{\eq}{\end{equation}}
\newcommand{\ve}{\varepsilon} \newcommand{\th}{\theta}
\newcommand{\la}{\lambda}
\newcommand{\thp}{\th+i\pi/2} \newcommand{\thm}{\th-i\pi/2}
\newcommand{\thpp}{\th+i\pi} \newcommand{\thmm}{\th-i\pi}
\newcommand{\iy}{\infty} \newcommand{\ov}{\over} \newcommand{\inv}{^{-1}}
\newcommand{\ra}{\rightarrow} \renewcommand{\sp}{\vspace{2ex}}
\newcommand{\noi}{\noindent} \newcommand{\tn}{\otimes} \def\Cm{{\bf
C}$\backslash0$}
\newcommand{\eth}{e^{\th}}\newcommand{\ethp}{e^{\th'}} \newcommand{\bR}{{\bf
R}}
\newcommand{\sech}{\mbox{sech}} \newcommand{\ph}{\varphi}
\newcommand{\iyy}{\int_{-\iy}^{\iy}} \newcommand{\dl}{\delta}
\newcommand{\Sp}{{\cal S}_{\pi}} \newcommand{\Spt}{{\cal S}_{\pi/2}}
\begin{center}{ \large\bf  Proofs of Two Conjectures Related to}\end{center}
\begin{center}{\large\bf the Thermodynamic Bethe Ansatz}\end{center}
\sp\begin{center}{{\bf Craig A. Tracy}\\
{\it Department of Mathematics and Institute of Theoretical Dynamics\\
University of California, Davis, CA 95616, USA\\
E-mail address: tracy@itd.ucdavis.edu}}\end{center}
\begin{center}{{\bf Harold Widom}\\
{\it Department of Mathematics\\
University of California, Santa Cruz, CA 95064, USA\\
E-mail address: widom@math.ucsc.edu}}\end{center}\sp

\renewcommand{\theequation}{1.\arabic{equation}}
\begin{abstract}
We prove that the solution to a pair of nonlinear integral equations
arising in the thermodynamic Bethe Ansatz can be expressed in terms
of the resolvent kernel of the linear integral operator with kernel
\[ {e^{-(u(\th)+u(\th'))}\ov\cosh{\th-\th'\ov2}}\, .\]
\end{abstract}
\noi{\bf I. Introduction}\sp

Thermodynamic Bethe Ansatz techniques were introduced in the pioneering
analysis of
Yang and Yang~\cite{yang} of the thermodyamics of a nonrelativistic,
one-dimensional Bose gas with delta function interaction.  Later this method
was extended to  a relativistic system with a
factorizable $S$-matrix to give an exact expression for the ground state
energy of this system on a cylindrical space of circumference
 $R$~\cite{klassen,Z1}.  This was
done by relating the ground state energy
 to the  free energy of the same system
 on an infinite line at temperature $T=1/R$.
In all  cases one expresses the various quantities of
interest in terms of ``excitation energies'' $\ve_a(\th)$ which are solutions
of nonlinear integral equations of the form
\[
\ve_a(\th)= u_a(\th) -\sum_b \int \phi_{ab}(\th-\th^\prime)
\log\left(1+z_b\, e^{-\ve_b(\th^\prime)}\right)\,{d\th^\prime\ov2\pi}\ \ \ \
(a=1,\,2,\,\cdots)\]
where $\phi_{ab}(\th)$ are expressible in terms of the 2-body $S$-matrix,
$z_a$ are  activities, and for relativistic systems
$u_a(\th)=m_a R \cosh\th$.  These nonlinear integral
equations  are the so-called
{\it thermodynamic Bethe Ansatz (TBA) equations\/}.
Solving the TBA equations is another matter.   The methods used are either
numerical or perturbative and there are, as far as the authors are aware, no
known
explicit  solutions to the TBA equations.
\par
It thus came as a surprise when
Cecotti et al.~\cite{cecotti2} (see also \cite{fendley}),
 in their analysis of certain $N=2$ supersymmetric
theories~\cite{cecotti1}, discovered  that a certain
quantity (the ``supersymmetric index''), expressible in terms of
the solution of the pair of ``TBA-like'' integral equations
\[
\ve(\th)=t\,\cosh\th-{1\ov2\pi}\int_{-\iy}^{\iy}{\log(1+\eta^2(\th'))\ov
\cosh(\th-\th')}\,d\th'\]
\[\eta(\th)=2\la\int_{-\iy}^{\iy}{e^{-\ve(\th')}\ov\cosh(\th-\th')}\,d\th' \,
,\]
 is also  expressible
in terms of a Painlev\'e III function with independent
variable $t$. (These TBA-like equations are a
perturbation of a system of TBA equations; see (5.7)--(6.8) in
\cite{cecotti2}.)\
Using results of McCoy et al.~\cite{mtw} on Painlev{\'e} III,  they expressed
this
supersymmetric index
as an infinite series related to the resolvent of the integral operator
with kernel
\[{e^{-{t\ov2}(\cosh\th+\cosh\th')}\ov\cosh{\th-\th'\ov2}}\> .\]
Zamolodchikov \cite{Z2} then conjectured that the system of
nonlinear equations could actually be
solved in terms of this resolvent kernel.
More precisely, if we denote the operator by $K$, the
kernel of the operator $K(I-\la^2K^2)\inv$ by $R_+(\th,\th')$, and set
$R_+(\th):=
R_+(\th,\th)$ then the system should be satisfied if
\[e^{-\ve(\th)}=R_+(\th)\]
and $\eta$ is defined by the second equation. In fact he conjectured that
this should hold for operators with kernels of the more general form
\bq {e^{-(u(\th)+u(\th'))}\ov\cosh{\th-\th'\ov2}}\label{K}\eq
if the first equation is replaced by
\[\ve(\th)=2\,u(\th)-{1\ov2\pi}\int_{-\iy}^{\iy}{\log(1+\eta^2(\th'))\ov
\cosh(\th-\th')}\,d\th'.\]
In addition he conjectured that, with the same function $\eta$,
\[R_-(\th)={1\ov\pi}R_+(\th)\,\int_{-\iy}^{\iy}{\arctan\eta(\th')\ov
\cosh^2(\th-\th')}\,d\th',\]
where $R_-(\th,\th')$ is the kernel of $K^2(I-\la^2K^2)\inv$ and $R_-(\th):=
R_-(\th,\th)$. (We state everything in terms of kernels here; in the cited work
the functions $R_{\pm}$ were given by infinite series. That they are the same
follows
from the Neumann series representation for the kernel of $(I-\la^2K^2)\inv$.)

We prove these conjectures here. One of the main ingredients is the fact that
the
equations are in a sense equivalent to relations among the analytic
continuations of
the functions $R_{\pm}(\th)$ into a strip. (See formula (6.8) of \cite{Z2}.)
Another is
a particularly convenient representations for these functions in terms of other
functions, which we call $Q(\th)$ and $P(\th)$. \
(That these latter functions are fundamental is known from earlier work
\cite{its90,tw5,tw5_5,tw7}.)\
These representations are stated
in Lemma 1, and in Lemmas 2 and 3 we state precise versions of the equivalence
alluded
to before. In Lemmas 4 and 5 we derive general properties of functions in the
range of
the operator $K$ in order to derive, as Lemma 6, some basic properties of the
functions
$Q(\th)$ and $P(\th)$.

If we try to prove the desired relation among the analytic continuations of
$R_{\pm}(\th)$ we find that we have to prove a certain crucial identity
involving
$Q(\th)$ and $P(\th)$ which is by no means obvious. But once conjectured it is
not hard
to prove, given the previous preparatory work, and that is stated as the
Proposition
which follows the lemmas. We show that a certain combination of these
functions,
which is clearly analytic in the strip, extends by periodicity to an entire
function.
Combining this fact with the use of Liouville's theorem, we deduce the
identity.

It should be mentioned that the main part of the argument may only
be used if $u$ belongs to a restricted class, but the result for general $u$
follows
by an approximation argument. This will be presented in the Appendix, as will
proofs
of some of the lemmas and some facts about Fourier transforms we shall use.
\sp

\setcounter{equation}{0}\renewcommand{\theequation}{2.\arabic{equation}}
\noi{\bf II. Preliminaries}\sp

We shall assume throughout that $u$ is continuous and bounded from below and
that
\bq 0<\la<e^{2\min u}/2\pi.\label{la}\eq
This assures that the series defining $R_{\pm}(\th)$ converge uniformly and
that
the operator $\la K$, acting on any of the usual function spaces, has norm less
than
1, so that $I-\la^2K^2$ is invertible. (This follows from (\ref{cosh}) below.)
Since the parameter $\la$
may be incorporated into $u$ we may assume that in fact $\la=1$. If we set
\bq E(\th):=\sqrt{2}\,e^{-u(\th)}\,e^{\th/2},\label{E}\eq
then the kernel of $K$ is given by
\bq {E(\th)E(\th')\ov \eth+\ethp}.\label{KE}\eq
Our functions $Q$ and $P$ are defined by
\[Q:=(I-K^2)\inv E,\quad P:=(I-K^2)\inv KE.\]

\noi{\bf Lemma 1}. We have the representations
\bq R_+(\th)={Q(\th)^2-P(\th)^2\ov 2\,e^{\th}},\quad R_-(\th)
={Q'(\th)\,P(\th)-P'(\th)\,Q(\th)\ov e^{\th}}.\label{Rreps}\eq

\noi{\bf Proof}. We use the notations $[A,\,B]:=AB-BA,\ \{A,\,B\}:=AB+BA$ and
write $X\tn Y$ for the operator with kernel $X(\th)\,Y(\th')$ and $M$ for
multiplication by $\eth$. Then we have immediately
\[\{M,\,K\}=E\tn E\]
from which it follows that also that
\[[M,\,K^2]=E\tn KE-KE\tn E\]
and then that
\[[M,\,(I-K^2)\inv]=Q\tn P-P\tn Q.\]
Since
\bq K^2(I-K^2)\inv=(I-K^2)\inv-I,\label{KI}\eq
we deduce that the kernel of this operator is given by the formula
\[R_-(\th,\th')={Q(\th)P(\th')-P(\th)Q(\th')\ov \eth-\ethp}.\]
The second part of (\ref{Rreps}) follows.

To derive the first part we use the general identity
\[\{A,\,BC\}=\{A,\,B\}\,C-B\,[A,\,C]\]
with $A=M,\ B=K$ and $C=(I-K^2)\inv$ together with the formulas above to deduce
\[\{M,\,K(I-K^2)\inv\}=E\tn Q-K(Q\tn P-P\tn Q).\]
Of course $KQ=P$, and applying (\ref{KI}) to $E$ gives
\[KP=Q-E,\]
and so the right side of the previous identity simplifies to $Q\tn Q-P\tn P$.
This
gives the representation for the kernel of $K(I-K^2)\inv$
\[R_+(\th,\th')={Q(\th)Q(\th')-P(\th)P(\th')\ov\eth+\ethp},\]
and the first part of (\ref{Rreps}) follows.\sp

Recall that a function $f$ defined on \bR\ is said to belong to the Wiener
space $W$
if its Fourier transform $\hat f$ belongs to $L_1$. Such a function is
necessarily
continuous and vanishes at $\pm\iy$. A sufficient condition that $f\in W$ is
that $f$ and
$f'$ belong to $L_2$. (See the Appendix.)

We use the notation ${\cal S}_{a}$ to denote the strip $|\Im\,\th|<a$ in the
complex
$\th$-plane, and $A({\cal S}_{a})$ to denote those functions $g$ which are
bounded and
analytic in the strip, continuous on its closure, and for which $g(\th+iy)\ra0$
as
$\th\ra+\iy$ through real values when $y\in\bR$ is fixed and satisfies
$|y|<a$.\sp

The proofs of the next three lemmas will be found in the Appendix.\sp

\noi{\bf Lemma 2}. Assume $f\in W$. If
\bq g(\th)={1\ov2\pi}\int_{-\iy}^{\iy}{f(\th')\ov\cosh(\th-\th')}\,
d\th'\label{fgint}\eq
then $g\in A(\Spt)$ and its boundary functions satisfy
\bq g(\thp)+g(\thm)=f(\th)\label{fg}\eq
for real $\th$.
Conversely, if $g\in A(\Spt)$  and if (\ref{fg}) holds then so does
(\ref{fgint}).\sp

\noi{\bf Lemma 3}. Assume $f,\,f'\in W$. If
\[ g(\th)={1\ov2\pi}\int_{-\iy}^{\iy}{f(\th')\ov\cosh^2(\th-\th')}\,d\th'\]
then $g\in A(\Spt)$ and its boundary functions satisfy
\[ g(\thp)-g(\thm)=if'(\th)\]
for real $\th$.
Conversely, if $g\in A(\Spt)$ and the second relation holds then so does the
first.\sp

\noi{\bf Lemma 4}. Assume $f\in L_1$ and set $h(x)$ equal to
$f(\log x)/x$ for $x>0$ and equal to 0 for $x\leq0$. If $h\in W$ and
\[g(\th)=\iyy{f(\th')\ov \eth+\ethp}\,d\th'\]
then $g\in A(\Sp).$\sp

These are the basic ingredients we shall use. We specialize at first to the
case where
$u$ is a Laurent polynomial in $\eth$,
\bq u(\th)=\sum_{k=-m}^n a_k\,e^{k\th}\label{u}\eq
for some $m,\,n>0$ with $a_{-m},\,a_n>0$. Thus $u(\th)$ is an entire function
of period $2\pi i$, and $E(\th)$ and its derivative tend exponentially to
0 as $\th\ra\pm\iy$ through real values. It follows from this that if
$f\in L_2$ then $Kf$ and its derivative are exponentially small
at $\pm\iy$. In particular, any function in the range of $K$ satisfies the
hypothesis of the next lemma.\sp

\noi{\bf Lemma 5}. If $f$ is a bounded function on $\bR$ with bounded
derivative then
$Kf(\th)/E(\th)$ extends to a function in $A(\Sp)$. The
boundary functions of $Kf$ satisfy
\[Kf\,(\thpp)+Kf\,(\thmm)=4\pi v(\th)f(\th)\]
for real $\th$, where
\[ v(\th):=e^{-(u(\th)+u(\thpp))}.\]

\noi{\bf Proof}. If we look at the the expression (\ref{KE}) for the kernel of
$K$
we see that $Kf(\th)/E(\th)$ is of the form of the function $g$ of Lemma 4 if
$f(\th)$ there is replaced by our $E(\th)f(\th)$. It is easy to see that if our
$f$
satisfies the stated conditions then the function $h$ in the statement of Lemma
4
belongs to $L_2$ and has an $L_2$ derivative, so $h\in W$ and the conclusion of
the
Lemma holds. Thus $Kf/E\in A(\Sp)$. For the boundary function identity
we use the expression (\ref{K}) for the kernel of $K$. If we
make the substitutions $\th\ra2\th,\ \th'\ra2\th'$ we see that
$e^{u(2\th)}\,(Kf)(2\th)=2\,\eth\,(Kf)(2\th)/E(2\th)$ is
exactly of the form of the function $g$ in the statement of Lemma 2 if the
function
$f(\th)$ there is replaced by our present $4\pi e^{-u(2\th)}\,f(2\th)$.
Applying the
identity stated there and using the periodicity of $u$ give the identity stated
here.\sp

We apply this to the functions $Q(\th)$ and $P(\th)$.\sp

\noi{\bf Lemma 6}. The functions
\bq{Q(\th)\ov E(\th)}-1\ \mbox{and}\ {P(\th)\ov E(\th)}\label{QPE}\eq
belong to $A(\Sp)$ and the boundary functions of $Q$ and $P$ satisfy the
identities
\bq Q(\thpp)+Q(\thmm)=4\pi v(\th)P(\th),\quad P(\thpp)+P(\thmm)=4\pi
v(\th)Q(\th)\label{QP}\eq
for real $\th$.\sp

\noi{\bf Proof}. Because of the relations $Q=E+K^2(I-K^2)\inv E,\ P=KQ$ all
statements
of the lemma except for the first part of (\ref{QP}) follow from Lemma 5 and
the
remark preceding it. Since
$Q-E=KP$ Lemma~5 gives the identity
\[(Q-E)(\thpp)+(Q-E)(\thmm)=4\pi v(\th)P(\th).\]
But it follows from the definition (\ref{E}) of $E$ and the fact that $u$ has
period
$2\pi i$ that\linebreak $E(\thpp)+E(\thmm)=0$. Thus we obtain the desired
identity
for $Q$.\sp

Using Lemma 6 and the fact that $v$ is an entire function we can conclude that
$Q$
and $P$ have analytic continuations to entire functions of $\th$; each use of
the
pair of identities allows us to widen by $\pi$ the strip of analyticity. Here
is the
crucial identity relating these continuations from which our results will
follow.\sp

\noi{\bf Proposition}. We have
\[Q(\thp)Q(\thm)-P(\thp)P(\thm)\]\[=E(\thp)\,E(\thm).\label{QQid}\]
{\bf Proof}. Set
\[S(\th):=Q(\thp)\,Q(\thm)-P(\thp)\,P(\thm).\]
Then
\[S(\thp)=Q(\thpp)\,Q(\th)-P(\thpp)\,P(\th),\]
\[S(\thm)=Q(\thmm)\,Q(\th)-P(\thmm)\,P(\th),\]
and so by (\ref{QP}),
\[S(\thp)+S(\thm)=4\pi v(\th)[P(\th)\,Q(\th)-Q(\th)\,P(\th)]=0.\]
It follows that $S(\th)$ extends to an entire function of period $2\pi i$ whose
values
at $\th\pm i\pi/2$ are negatives of each other. Therefore
$1-S(\th)/E(\thp)\,E(\thm)$
extends to an entire function of period $\pi i$ whose values at $\th\pm i\pi/2$
are
equal. (We used again the fact that
$E(\thmm)=-E(\thpp)$.) To show that this is 0 (this is equivalent to the
claimed identity) it suffices, by Liouville's theorem, to show that it is
bounded and
that it tends to 0 as $\th\ra+\iy$ through real values. For a $\pi i$-periodic
function it suffices to show that these properties hold in the strip $\Spt$.
They do hold there because for $\th$ in this strip $\th\pm i\pi/2$ lie in the
strip
$\Sp$, for which we have the conclusions of Lemma 6.\sp

\setcounter{equation}{0}\renewcommand{\theequation}{3.\arabic{equation}}
\noi{\bf III. Proof of the Conjectures}\sp

We have to show
that if $\ve:=-\log\,R_+$ and if $\eta$ is defined by
\bq\eta(\th)=2\int_{-\iy}^{\iy}{e^{-\ve(\th')}\ov\cosh(\th-\th')}\,d\th'
\label{TBA1}\eq
(recall that we have taken $\la=1$), then we have the two identities
\bq 2\,u(\th)-\ve(\th)={1\ov2\pi}\int_{-\iy}^{\iy}{\log(1+\eta^2(\th'))\ov
\cosh(\th-\th')}\,d\th',\label{TBA2}\eq
\bq R_-(\th)={1\ov\pi}R_+(\th)\,\int_{-\iy}^{\iy}{\arctan\eta(\th')\ov
\cosh^2(\th-\th')}\,d\th'.\label{TBA3}\eq

We shall assume that we have a function $\eta$ satisfying
(\ref{TBA1})--(\ref{TBA3}),
and formally apply the first parts of Lemmas 2 and 3 to obtain the three
identities
(\ref{TBA1.1})--(\ref{TBA3.1}) below. From these we shall see that $\eta$ must
have a
certain representation in terms of $Q$ and $P$. Then, using the identities of
the
Proposition, we shall show that (\ref{TBA1.1})--(\ref{TBA3.1}) in fact hold if
$\eta$
is defined this way. Finally, using the second parts of Lemmas 2 and 3, we show
that
(\ref{TBA1})--(\ref{TBA3}) hold. (Of course the first step is unnecessary for
the
proof, but it provides motivation for the eventual definition of $\eta$.)

Applying Lemmas 2 and 3 to (\ref{TBA1})--(\ref{TBA3}) give
\bq 4\pi\,R_+(\th)=\eta(\thp)+\eta(\thm),\label{TBA1.1}\eq
\bq \log(1+\eta^2(\th))=2u(\thp)-\ve(\thp)+2u(\thm)-\ve(\thm),\label{TBA2.1}\eq
\bq 2i{\eta'(\th)\ov1+\eta(\th)^2}={R_-(\thp)\ov R_+(\thp)}-{R_-(\thm)\ov
R_+(\thm)}.
\label{TBA3.1}\eq
Exponentiating (\ref{TBA2.1}) and using the definition of $\ve$ give
\bq 1+\eta^2(\th)=
R_+(\thp)\,R_+(\thm)\,e^{2\,(u(\thp)+u(\thm))}.\label{TBA2.2}\eq
By Lemma 1 and (\ref{E}) this may be written
\bq 1+\eta^2(\th)={(Q^2-P^2)(\thp)\cdot(Q^2-P^2)(\thm)\ov E(\thp)^2E(\thm)^2}.
\label{TBA2.3}\eq
Lemma 1 shows that (\ref{TBA3.1}) may be written
\[i{\eta'(\th)\ov1+\eta^2(\th)}={Q'P-P'Q\ov Q^2-P^2}(\thp)-{Q'P-P'Q\ov
Q^2-P^2}(\thm).\]
Taking the logarithmic derivative of both sides of (\ref{TBA2.3}) and dividing
by 2 gives
\[\eta(\th){\eta'(\th)\ov1+\eta^2(\th)}={Q'Q-P'P\ov Q^2-P^2}(\thp)+
{Q'Q-P'P\ov Q^2-P^2}(\thm)-{E'\ov E}(\thp)-{E'\ov E}(\thm).\]

We have a choice now of either adding or subtracting the last two displayed
formulas.
Choosing the former, we obtain
\[{\eta'(\th)\ov\eta(\th)-i}={Q'-P'\ov Q-P}(\thp)+{Q'+P'\ov Q+P}(\thm)-
{E'\ov E}(\thp)-{E'\ov E}(\thm),\]
or equivalently
\[{\eta'(\th)\ov\eta(\th)-i}={d\ov d\th}\log(Q-P)(\thp)+{d\ov
d\th}\log(Q+P)(\thm)-
{d\ov d\th}\log E(\thp)E(\thm).\]
Integrating and exponentiating gives the desired formula
\bq \eta(\th)-i=-i{(Q-P)(\thp)\cdot(Q+P)(\thm)\ov
E(\thp)E(\thm)}.\label{eta1}\eq
The reason the constant factor on the right must be $-i$ is that we want
$\eta(+\iy)
=0$. (See (\ref{TBA2.3}) and Lemma 6.)
If we had subtracted before instead of adding we would have been led to the
similar
but apparently different formula
\bq \eta(\th)+i=i{(Q+P)(\thp)\cdot(Q-P)(\thm)\ov
E(\thp)E(\thm)}.\label{eta2}\eq

We shall now show that if $\eta$ is defined by (\ref{eta1}) then (\ref{eta2})
also holds, as do relations (\ref{TBA1.1})--(\ref{TBA3.1}).

First, the statement that the right side of (\ref{eta2}) minus the right side
of
(\ref{eta1}) is equal to $2i$ follows from the Proposition. Thus (\ref{eta1})
and
(\ref{eta2}) are completely equivalent.

Second, taking the product of (\ref{eta1}) and (\ref{eta2}) gives
(\ref{TBA2.3}) and
hence (\ref{TBA2.1}).

Third, reversing the argument that showed  (\ref{TBA2.3}) and (\ref{TBA3.1})
imply (\ref{eta1}) we see that (\ref{eta1}) and (\ref{TBA2.3}), which we now
have,
imply (\ref{TBA3.1}).

Finally, to obtain (\ref{TBA1.1}) we use (\ref{eta2}) to express $\eta(\thp)$
and
(\ref{eta1}) to express\linebreak $\eta(\thm)$ and find that their sum equals
\[i{(Q-P)(\th)\;\big[(Q+P)(\thpp)+(Q+P)(\thmm)\big]\ov E(\th)E(\thpp)}.\]
(We used yet again the fact that $E(\thmm)=-E(\thpp)$.) By (\ref{QP}) this is
equal to
\[{4\pi i\,v(\th)\ov E(\th)E(\thpp)}[Q(\th)^2-P(\th)^2]\]
and by (\ref{Rreps}) and the definitions of $E(\th)$ and $v(\th)$ this equals
$4\pi R_+(\th)$.\sp

So (\ref{TBA1.1})--(\ref{TBA3.1}) are established. Now we show that they imply
(\ref{TBA1})--(\ref{TBA3}).
By Lemmas 2 and 3 this will be true if the functions
\bq R_+,\ \log(1+\eta^2),\ \arctan\,\eta,\ \eta'/(1+\eta^2)\label{L}\eq
belong to $W$ and the functions
\bq\eta,\ 2u-\ve,\ R_-/R_+\label{H}\eq
belong to $A(\Spt)$.

Our assumption has been that $\la=1$ satisfies inequality (\ref{la}).
This identity is still satisfied if $u$ is increased or, equivalently, if $E$
is
decreased. It follows that we could, in our representation (\ref{KE}) of the
kernel
of $K$, have replaced $E$ by $\dl E$ for any $\dl\in [0,\,1]$.
It is clear that all quantites in (\ref{TBA1})--(\ref{TBA3}) would then be
real-analytic functions of $\dl$ for $\dl\in [0,\,1]$. So if the relations hold
for
sufficiently small $\dl$ they hold for all, including $\dl=1$. If we retrace
the
steps leading to bounds for the functions (\ref{QPE}) we find that they tend to
0 as
$\dl\ra0$. (In fact, they are $O(\dl^2)$.) Hence we may assume that the bounds
for
these functions are as small as we like. In fact we assume that
\bq |{Q(\th)\ov E(\th)}-1|<{1\ov4}\ \mbox{and}
\ |{P(\th)\ov E(\th)}|<{1\ov4}\qquad(\th\in\Sp).\label{QPest}\eq

We take in succession the items we have to verify. Recall that a function
belongs to $W$
if it and its derivative belong to $L_2$. This is what we shall show for the
functions
(\ref{L}). First, from (\ref{E}) and
(\ref{Rreps}) we see have
\[ R_+(\th)={Q(\th)^2-P(\th)^2\ov E(\th)^2}e^{-2u(\th)}.\]
This first factor is bounded and analytic in $\Sp$ and so is bounded and has
bounded
derivative on \bR. The second factor is in $L_2$ and so is its derivative.
Hence $R_+\in W$.

The next three are not obvious. If we use the representation (\ref{KE}) for the
kernel of $K$ we see that for any $f$ we have
\[{Kf\ov E}(\thp)-{Kf\ov E}(\thm)=-2i\,\iyy{E(\th')f(\th')\eth\ov
e^{2\th}+e^{2\th'}}
\,d\th'=O(\sech\,\th)\]
if, say,  $f\in L_{\iy}$. This holds for $|\Im\,\th|$
strictly less than $\pi/4$. Applying this to $f=Q$ and $f=P$ we deduce that
\[{P\ov E}(\thp)-{P\ov E}(\thm)=O(\sech\,\th),\quad
{Q\ov E}(\thp)-{Q\ov E}(\thm)=O(\sech\,\th)\]
as long as $|\Im\,\th|<\pi/8$, for example. Now adding (\ref{eta1}) and
(\ref{eta2}) gives the
representation
\bq \eta(\th)=i\,{Q(\thm)\,P(\thp)-Q(\thp)\,P(\thm)\ov
E(\thm)E(\thp)}.\label{eta3}\eq
By the above and (\ref{QPest}), this is
\[\Big[{Q\ov E}(\thp)+O(\sech\,\th)\Big]\,{P\ov E}(\thp)-{Q\ov
E}(\thp)\,\Big[{P\ov
E}(\thp)+
O(\sech\,\th)\Big]=O(\sech\,\th).\]
(Observe that if $\th\in{\cal S}_{\pi/8}$ then $\th\pm i\pi/2\in\Sp$.) Since
this bound
on $\eta$ holds in a full strip, the same bound (with different constants)
holds
on \bR\ for each of the derivatives of $\eta$. Since $\eta$ is real-valued on
\bR\
(this follows from (\ref{eta3}) and the fact that $Q,\ P$, and $E$ are
real-valud on
\bR) we deduce that the last three functions in (\ref{L}), and all their
derivatives, belong to $L_2(\bR)$. Hence they all belong to $W$.

Now we show that the functions (\ref{L})
all belong to $A(\Spt)$. First, by (\ref{Rreps}),
\[{R_-\ov R_+}=2{Q'\,P-P'\,Q\ov Q^2-P^2}=2{Q'\,P-P'\,Q\ov Q^2}\,
{1\ov1-(P/Q)^2}.\]
By (\ref{QPest}) $P/Q$ is bounded in the larger strip $\Sp$ by 1/3 and so
it follows that the last factor above is bounded. It also follows that $P/Q\in
A(\Sp)$
from which it follows that $(P/Q)'\in A(\Spt)$. Hence $R_-/R_+\in A(\Spt)$.

Next, we use (\ref{Rreps}) again to write
\[2u-\ve=2u+\log{Q^2-P^2\ov2 \eth}=\log{Q^2-P^2\ov E^2}.\]
By (\ref{QPest}) we have $|1-(Q\pm P)/E|<1/2$ in $\Spt$ and we deduce as above
that
$2u-\ve\in A(\Spt)$.

Finally since $1-Q/E$ and $P/E$ belong to $A(\Sp)$, the functions
\[1-{Q(\th\pm i\pi/2)\ov E(\th\pm i\pi/2)},\quad {P(\th\pm i\pi/2)\ov E(\th\pm
i\pi/2)}\]
belong to $A(\Spt)$. It follows from this and the representation
(\ref{eta3}) that $\eta\in A(\Spt)$.\sp

Thus we have proved the conjectures in the case where $u(\th)$ has the special
form
(\ref{u}). For a general $u$, more precisely for any $u$ which is continuous
and
bounded below, we can find a sequence of $u_n$ of the special type such
that $e^{-u_n}$ converge boundedly and locally uniformly to $e^{-u}$. (This
will be
demonstrated in the Appendix.) This is enough
to deduce the result for $u$ from the results for the $u_n$.\sp

\setcounter{equation}{0}\renewcommand{\theequation}{4.\arabic{equation}}
\noi{\bf IV. Appendix}\sp

We give details here of certain matters postponed from the previous sections.
First we
recall some facts about the Fourier transform, which we denote, as usual, by a
circumflex:
\[\hat f(\xi)=\iyy e^{-i\xi\th}\,f(\th)\,d\th.\]
If $\hat f\in L_1$, in other words, if $f\in W$, we have the Fourier inversion
formula
\[f(\th)={1\ov 2\pi}\iyy e^{i\xi\th}\,\hat f(\xi)\,d\xi,\]
and $f(\pm\iy)=0$. If $||\ \ ||_p$ denotes the norm in the space
$L_p$ then by the inversion formula we have the inequality
\bq ||f||_{\iy}\leq{1\ov2\pi}||\hat f||_1.\label{1est}\eq
Parseval's identity reads
\[ ||\hat f||_2={1\ov\sqrt{2\pi}}||f||_2,\]
and we have the general formula $\widehat{f'}(\xi)=i\xi\hat f(\xi)$.\sp

\noi{\bf Proof that} {\boldmath $f,\ f'\in L_2$} {\bf implies} {\boldmath $f\in
W$}.
It suffices to show that if we
write
$\hat f(\xi)$ as \linebreak $[\hat f(\xi)\,(\xi+i)]\,(\xi+i)\inv$
then both factors on the right belong to $L_2$. The second factor surely does,
and the
square of the absolute value of the first factor equals
\[|\hat f(\xi)|^2\,(\xi^2+1)=|\widehat{f'}(\xi)|^2+|\hat f(\xi)|^2.\]
The right side belongs to $L_1$ by Parseval's identity and the assumption
$f,\ f'\in L_2$.\sp

\noi{\bf Proof of Lemma 2}. It is clear that $g$, when defined by
(\ref{fgint}), is
analytic in the strip. If we write $g_y(\th):=g(\th+iy)$ for real $\th$ then
\bq\widehat{g_y}(\xi)=\hat f(\xi)\,{e^{-y\xi}\ov2\,\cosh\pi\xi/2}.\label{gy}\eq
The first factor belongs to $L_1$, by assumption, and the second factor is
bounded by 1 for $|y|<\pi/2$. In particular $g_y\in W$ for each $y$ so
$g(\pm\iy+iy)=0$. From
\[\iyy|\widehat{g_y}(\xi)|\,d\xi\leq\iyy|\hat f(\xi)|\,d\xi\]
and (\ref{1est}) we see that $g_y$ is uniformly bounded for $|y|<\pi/2$.
Finally, as $y\ra\pm\pi/2$ the second factor in (\ref{gy}) converges pointwise
and boundedly. This implies (by the Lebesgue dominated convergence theorem)
that $\widehat{g_y}(\xi)$  converges in $L_1$ and so (by (\ref{1est}) again)
$g_y(\th)$ converges uniformly in $\th$. This implies that $g(\th)$ extends
continuously
to the closure of $\Spt$. Thus $g\in A(\Spt)$. The sum of the two limiting
values of
the second factor in (\ref{gy}) is equal to 1, and this gives (\ref{fg}).

To prove the converse, let $h$ denote the
difference between the two sides of (\ref{fgint}). Then, using our assumption
and what
we have already shown, $h\in A(\Spt)$ and its boundary functions satisfy
\[ h(\thp)+h(\thm)=0\]
for real $\th$. It follows that $h$ extends to a $2\pi i$-periodic entire
function.
This function is bounded, and so must be a constant, and the constant must be 0
because $h(+\iy)=0$.\sp

\noi{\bf Proof of Lemma 3}. In this case the Fourier transform of $g_y$ equals
\[\hat f(\xi)\,\xi\,{e^{-y\xi}\ov2\,\sinh\pi\xi/2}=\hat f(\xi)\,(\xi+i)\,
{\xi\,e^{-y\xi}\ov2\,(\xi+i)\,\sinh\pi\xi/2}.\]
Now $\hat f(\xi)(\xi+i)=-i\widehat{f'}(\xi)+i\hat f(\xi)\in L_1$, the last
factor on the
right is uniformly bounded
for $|y|<\pi/2$ and it converges pointwise as $y\ra\pm\pi/2$. We deduce as
before that
$g\in A(\Spt)$. The difference of the limiting values of the
last factor on the left side at $y=\pm\pi/2$ is $-1$ and so the difference of
the limiting values of $g$
has Fourier transform $-\hat f(\xi)\xi=\widehat{if'}$. This establishes the
first
part of the lemma and the second follows just as before.\sp

\noi{\bf Proof of Lemma 4}. For fixed $\th'$ the factor $(\eth+\ethp)\inv$ in
the
integral defining $g(\th)$ is analytic in $\th$. For $\th$ in any subset of
$\Sp$ of
the form
\[\Re\,\th\geq\th_0,\ \ \ |\Im\,\th|\leq\pi-\dl\ \,(\dl>0)\]
this factor is bounded uniformly in $\th'$ and tends to 0 pointwise as
$\Re\,\th\ra
+\iy$. Since $f\in L_1$ this is enough to conclude that $g$ is analytic in
$\Sp$ and
tends to 0 in this strip as $\Re\,\th\ra+\iy$ and $\Im\,\th$ is fixed.

It remains to prove boundedness of $g$ and continuity near the boundary of
$\Sp$, and
for this we use the function $h$. In the lower part of the strip, $-\pi<y<0$,
we set
$z=-e^{\th+iy}$ with $\th$ real, so that $\Im\,z>0$, and we can write
\[g(\th+iy)=\int_0^{\iy}{h(x)\ov x-z}\,dx=\iyy{h(x)\ov x-z}\,dx.\]
Using the Fourier inversion formula and interchanging the order of integration,
we
find the representation
\[g(\th+iy)=i\int_0^{\iy}e^{iz\xi}\hat h(\xi)\,d\xi.\]
Since $\hat h\in L_1$ the integral is bounded uniformly for $\Im\,z>0$ and we
deduce
as in the Lemmas 2 and 3 that it extends continuously to $\Im\,z=0$. Thus
$g(\th+iy)$
is bounded for $0\leq y<\pi$ and extends continuously to $y=\pi$.
A similar argument holds for the upper half of $\Sp$, and so $g\in A(\Sp)$.\sp

\noi{\bf Extension to general}\ {\boldmath $u$}. We begin with an approximation
fact,
reminiscent
of the Weierstrass approximation theorem. Recall the notation $C_0(\bR)$ for
the space
 of continuous
functions $f$ on \bR\ satisfying $f(\pm\iy)=0$, which is a Banach space under
the norm
\[||f||:=\sup\{|f(x)|:x\in\bR\}.\]
The fact is that for each $\dl>0$ the finite linear combinations
of the functions
\[\sinh^k\th\,e^{-\dl\,\sinh^2\th}\qquad(k=0,\,1,\,\cdots)\]
are dense in $C_0(\bR)$; in other words, for any $f\in C_0(\bR)$ and any
$\dl'>0$
there exists a finite linear combination
\[p(\th)=\sum_{k=0}^N a_k\,\sinh^k\th\]
such that
\bq |p(\th)\,e^{-\dl\,\sinh^2\th}-f(\th)|<\dl'\label{app}\eq
for all $\th$. This is true because the change of variable $t=\sinh\th$
converts it to the statement that the finite linear combinations of the
functions
$t^k\,e^{-\dl t^2}$ are dense in $C_0(\bR)$. And this in turn is
true because if it weren't then the Hahn-Banach theorem and Riesz
representation
theorem (\cite{R}, Thms. 5.19 and 6.19) would imply that there there is a
function of
bounded variation (= signed measure) $\mu$ on \bR, not identically zero, such
that
$\iyy t^k\,e^{-\dl t^2}\,d\mu(t)=0$
for all $k\geq0$. But then the entire function
$F(z):=\iyy e^{izt}\,e^{-\dl t^2}\,d\mu(t)$
would satisfy $F^{(k)}(0)=0$ for $k\geq0$ and so $F\equiv0$. This in turn
implies
$e^{-\dl t^2}\,d\mu(t)\equiv0$ and so $\mu(t)\equiv0$, a contradiction.

Here is how to construct the sequance $u_n$ described
at the end of Section III. We may clearly assume that $u$ is uniformly
positive,
i.e. that for some $\alpha>0$ we have $u(\th)\geq\alpha$ for all $\th$. Let $n$
be given
and define
\[w:=\min(u,\,n).\]
Then find $p(\th)$, a linear combination of the powers of $\sinh\,\th$, such
that
(\ref{app}) holds with
\[f(\th)=\sqrt{w(\th)}\,e^{-n\inv\sinh^2\th}\]
and $\dl=\dl'=n\inv$. The inequality may be rewritten
\[|p(\th)-\sqrt{w(\th)}|<n\inv\,e^{n\inv\sinh^2\th}.\]
It is an easy exercise to deduce from this that for sufficiently large $n$
\[|p(\th)^2-u(\th)|<6\,n^{-1/2}\quad{\rm if}\ u(\th)<n\ {\rm and}\
\sinh^2\th<n.\]
(We use here the facts that $u$ is uniformly positive and that $e<3$.) The
function
$p(\th)^2$ is our $u_n(\th)$.

We now deduce the identities (\ref{TBA1})--(\ref{TBA3}) for general $u$. Denote
by
$R_{n\pm}$ the $R_{\pm}$ functions associated with the functions $u_n$.
If we can show that $R_{n\pm}(\th)\ra R_{\pm}(\th)$ boundedly and pointwise
then (\ref{TBA1})--(\ref{TBA3}) for $u$ will follow from the corresponding
identities
for the $u_n$, since by the dominated convergence theorem we could take the
limits
as $n\ra\iy$ under the integral signs. The function
$R_+(\th)$ is given by the series
\bq \sum_{m=0}^{\iy}\la^{2m}\iyy\cdots\iyy K(\th,\,\th_1)\cdots
K(\th_{2m},\,\th)\,
d\th_1\cdots d\th_{2m}\label{R}\eq
where $K(\th,\,\th')$ is given by (\ref{K}). It follows from the fact
\bq \iyy{d\th\ov\cosh\,\th/2}=2\pi\label{cosh}\eq
that the series converges uniformly in $\th$ when $\la$ satisfies (\ref{la}).
By a real-analyticity argument already used
we may assume that $u$ is uniformly positive, so that by the
previous construction $u_n\geq0$ fo all $n$, and that
$\la<1/2\pi$. Denote by
$K_n(\th,\,\th')$
the kernel corresponding to $u_n$ so that $R_{n+}(\th)$ is given by the series
\bq\sum_{m=0}^{\iy}\la^{2m}\iyy\cdots\iyy K_n(\th,\,\th_1)\cdots
K_n(\th_{2m},\,\th)\,
d\th_1\cdots d\th_{2m}.\label{Rn}\eq
It follows from (\ref{cosh}) and the inequality $e^{-u_n}\leq1$ that the
integral in
the $m$th term of (\ref{Rn}) is at most $(2\pi)^{2m}$ for all $n$ and so,
since $\la<1/2\pi$, the series converges uniformly in $n$. Thus we may take
the limit as $n\ra\iy$ under the summation sign. Next, each integrand
$K_n(\th,\,\th_1)\cdots K_n(\th_{2m},\,\th)$ is uniformly bounded by
$K(\th,\,\th_1)\cdots K(\th_{2m},\,\th)$, which has finite integral over
$\bR^{2m}$,
and so we may take each limit as $n\ra\iy$ under the integral sign (again by
the
dominated convergence theorem). The result
is the series (\ref{R}), and this gives
\[\lim_{n\ra\iy}R_{n+}(\th)= R_+(\th).\]
Since $0<R_{n+}(\th)\leq R_+(\th)$ we have established that $R_{n+}(\th)\ra
R_+(\th)$
boundedly and pointwise. A similar argument applies to $R_{n-}(\th)$, and the
proof
is complete.\sp

\par
\begin{center}{\bf Acknowledgements}\end{center}
The authors wish to thank Paul Fendley for elucidating the conjectural status
of the
identities we prove here. This work was supported in part by
the National Science Foundation through grants DMS--9303413 and DMS--9424292.


\newpage
\begin{thebibliography}{4}

\bibitem{cecotti1} Cecotti, S., Vafa, C.: {\sl Topological--anti-topological
fusion\/},
Nucl.\ Phys.\ {\bf B367}, 359--461 (1991); {\sl Ising model and $N=2$
supersymmetric
theories\/}, Commun.\ Math.\ Phys.\ {\bf 157}, 139--178 (1993)

\bibitem{cecotti2} Cecotti, S., Fendley, P., Intriligator, K., Vafa, C.:
{\sl A new supersymmetric index\/}, Nucl.\ Phys.\ {\bf B386}, 405--452 (1992)

\bibitem{fendley} Fendley, P., Saleur, H.: {\sl $N=2$ supersymmetry,
Painlev{\'e} III and exact scaling functions in 2D polymers\/},
Nucl.\ Phys.\ {\bf B388}, 609--626 (1992)

\bibitem{its90} Its, A.~R.,  Izergin, A.~G.,  Korepin, V.~E.,
Slavnov, N.~A.:  {\sl Differential equations for quantum
correlation functions\/}, Int.\ J.\ Mod.\ Physics B {\bf 4},
1003--1037 (1990)

\bibitem{klassen} Klassen, T.~R., Melzer, E.: {\sl The thermodynamics of
purely elastic scattering theories and conformal perturbation theory\/},
Nucl.\ Phys.\ {\bf B350}, 635--689 (1991)

\bibitem{mtw} McCoy, B.~M., Tracy, C.~A., Wu, T.~T.:
{\sl Painlev{\'e} functions of the third kind\/}, J.\ Math.\ Phys.\
{\bf 18}, 1058--1092 (1977)

\bibitem{R} Rudin, W.: {\sl Real and Complex Analysis, 3rd ed\/}. New York:
McGraw-Hill, 1987

\bibitem{tw5} Tracy, C.~A.,  Widom, H.: {\sl Fredholm determinants,
differential
equations and matrix models\/}.  Commun.\  Math.\  Phys.\  {\bf 163},  33--72
(1994)

\bibitem{tw5_5} Tracy, C.~A.,  Widom, H.: {\sl Systems of partial differential
equations
for a class of operator determinants\/}.  Operator Theory: Adv.\ and Appls.\
{\bf 78}, 381--388 (1995)

\bibitem{tw7} Tracy, C.~A.,  Widom, H.: {\sl Fredholm determinants and the
mKdV/sinh-Gordon hierarchies\/}, to appear in Commun.\  Math.\  Phys.,
solv-int/9506006

\bibitem{yang} Yang, C.~N., Yang, C.~P.: {\sl Thermodynamics of a
one-dimensional
system of bosons with repulsive delta-function interaction\/}, J.\ Math.\
Phys.\
{\bf 10}, 1115--1122 (1969)

\bibitem{Z1} Zamolodchikov, Al.~B.:{\sl Thermodynamic
Bethe Ansatz in relativistic models: Scaling 3-state Potts
and Lee-Yang models\/}, Nucl.\ Phys.\ {\bf B342}, 695--720 (1990)

\bibitem{Z2} Zamolodchikov, Al.~B.:{\sl  Painlev{\'e} III and 2D polymers\/}.
Nucl.\ Phys.\ {\bf B432}[FS], 427--456 (1994)

\end{thebibliography}
\end{document}